\documentclass{article}

\usepackage[svgnames, x11names]{xcolor}

\newif\ifeurocr 

\eurocrtrue 

\ifeurocr
\newcommand{\gf}[2]{#2}
\else 
\newcommand{\gf}[2]{#1}
\fi

\usepackage{listings}

\usepackage[pdftex]{graphicx}
\graphicspath{{./figures/}}
\DeclareGraphicsExtensions{.pdf}

\usepackage[cmex10]{amsmath}
\usepackage{amssymb}
\usepackage{mathtools}

\usepackage[caption=false]{subfig}

\usepackage{hyperref} 

\usepackage{xspace}
\usepackage[nocompress]{cite}

\usepackage{enumitem}

\usepackage[english]{babel} 

\newcommand{\vlet}{VIoLET\xspace}

\begin{document}

\title{VIoLET: A Large-scale Virtual Environment for Internet of Things~\thanks{To appear in the Proceedings of the 24TH INTERNATIONAL EUROPEAN CONFERENCE ON PARALLEL AND DISTRIBUTED COMPUTING (EURO-PAR), August 27--31, 2018, Turin, Italy, \href{https://europar2018.org/program}{europar2018.org}}~\thanks{Selected as a \emph{Distinguished Paper} for presentation at the Plenary Session of the conference}}
%
%
\author{Shreyas Badiger, Shrey Baheti and Yogesh Simmhan\\
		\normalsize{\emph{Department of Computational and Data Sciences}}\\
		\normalsize{\emph{Indian Institute of Science (IISc), Bangalore 560012, India}}\\
		\normalsize{\emph{Email: shreyasb@IISc.ac.in, shreybaheti@IISc.ac.in, simmhan@IISc.ac.in}}}
	
	\date{}

\maketitle              

\begin{abstract}
IoT deployments have been growing manifold, encompassing sensors, networks, edge, fog and cloud resources. Despite the intense interest from researchers and practitioners, most do not have access to large-scale IoT testbeds for validation. Simulation environments that allow analytical modeling are a poor substitute for evaluating software platforms or application workloads in realistic computing environments. Here, we propose \vlet, a virtual environment for defining and launching large-scale IoT deployments within cloud VMs. It offers a declarative model to specify container-based compute resources that match the performance of the native edge, fog and cloud devices using Docker. These can be inter-connected by complex topologies on which private/public networks, and bandwidth and latency rules are enforced. Users can configure synthetic sensors for data generation on these devices as well. We validate \vlet for deployments with $>400$ devices and $>1500$ device-cores, and show that the virtual IoT environment closely matches the expected compute and network performance at modest costs. This fills an important gap between IoT simulators and real deployments. 
\end{abstract}

\section{Introduction}
\label{sec:intro}
Internet of Things (IoT) is expanding rapidly as diverse domains deploy sensors, communication, and gateway infrastructure to support applications such as smart cities, personalized health, and autonomous vehicles. IoT is also accelerating the need for, and the use of edge, fog and cloud resources, in a coordinated manner. The need comes from the availability of large volumes of data streams that need to be analyzed closer to the edge to conserve bandwidth (e.g., video surveillance), or of fast data streams that need to be processed with low latency~\cite{satya:pervasive:2015}. Edge gateway devices such as Raspberry Pi and Smart Phones have non-trivial resource capabilities, and can run a full Linux stack on 64-bit ARM processors. Fog devices such as NVidia's TX1 and Dell's Edge Gateways have power-efficient Atom processors or GPUs to support the needs of several edge devices~\cite{bonomi2012fog,Varshney:2017}. At the same time, edge and even accelerated fog devices may not have the elastic and seemingly infinite on-demand resource capacity that is available in the cloud, and necessary for processing by certain IoT applications.

Besides production deployments of IoT, there is also active research at the intersection of IoT, and edge, fog and cloud computing that is investigating application scheduling, resiliency, big data platforms, and so on~\cite{dasterdi:corr:2016,ghosh:tcps:2017}. However, a key gap that exists is the ability to validate these research outcomes on real or realistic IoT environments. Research IoT testbeds may have just 10's of devices, and simulation environments make too many idealized assumptions and do not allow actual applications to be deployed. Manually launching and configuring containers is time consuming and error-prone. Even planning of production deployment of IoT, edge and fog resources are based on analytical models or simulations, which may not hold in practice~\cite{gupta:2016,sonmez:2017,leland:2017}. 

What is lacking is a virtualized IoT environment that offers the computing and network ecosystem of a real deployment without the need to purchase, configure and deploy the edge, fog and networking devices. Here, we propose \emph{VIoLET, a \underline{L}arge-scale \underline{V}irtual \underline{E}nvironment for \underline{I}nternet \underline{o}f \underline{T}hings}. \vlet offers several essential features that make it valuable for researchers and planners. It is a virtualized environment that uses containers to offer comparable compute resources as edge, fog and cloud, and can run real applications. It allows the easy definition of diverse network topologies, and imposes bandwidth and latency limits between containers. \vlet also allows the definition of virtual sensors that generate data with various distributions within the containers. It runs on top of cloud VMs or commodity clusters, allowing it to scale to hundreds or thousands of devices, provided cumulative compute capacity is available on the host machines. All of these help setup and validate an environment that mimics the behavior of city-scale IoT deployments in a fast, reproducible and cost-effective manner. \emph{\vlet v1.0} is available for download from \url{https://github.com/dream-lab/VIoLET}.

The rest of this paper is organized as follows. We motivate various requirements for \vlet in Section~\ref{sec:req}, describe its architecture design that meets these requirements and its implementation in Section~\ref{sec:arch}, present results on deploying and scaling \vlet for different IoT topologies in Section~\ref{sec:eval}, compare it with related literature and tools in Section~\ref{sec:related}, and finally present our conclusions and future work in Section~\ref{sec:conclusions}.

\section{Design Requirements}
\label{sec:req}
Here, we present high-level requirements for a \emph{Virtual Environment (VE)} like \vlet, based on the needs of researchers and developers of applications, platforms and runtime environments for IoT, edge, and fog resources.

\textbf{Compute environment.} The VE should provide the ability to configure computing resources that capture the performance behavior of \emph{heterogeneous IoT resources}, such as edge devices, gateways, fog and even cloud resources. Key resource capabilities to be controlled include CPU rating, memory and storage capacity, and network. Further, a \emph{compute environment} that can host platforms and run applications should be provided within these resources. Virtual Machines (VM) have traditionally offered such capabilities, but are too heavy-weight for the often light-weight and plentiful IoT devices. \emph{Containers} are much more light-weight and offer similar capabilities. One downside is the inability to change the underlying Operating System (OS) as it is coupled with the Linux kernel of the host machine. However, we expect most IoT devices to run a flavor of Linux.

\textbf{Networking.}
Communication is central to IoT, and the networking layer is sensitive to various deployment limitations on the field. Wired, wireless and cellular networks are common, each with different \emph{bandwidth and latency characteristics}. There is also a distinction between \emph{local and wide area networks}, and \emph{public and private networks} -- the latter can limit the visibility of devices to each other. These affect the platforms and applications in the computing environment, and can decide who can connect to whom and if an indirection service is required. The VE needs to capture such diverse network topologies and behavior.

\textbf{Sensing and Data Streams.}
Sensors (and actuators) form the third vital component of IoT. These are often connected to the edge computing devices by physical links, \emph{ad hoc} wireless networks, or even on-board the device. These form the source of the distributed, fast data streams that are intrinsic to IoT deployments. The VE should provide the ability to simulate the generation of sensor event streams with various sampling rates and distributions at the compute devices for consumption by hosted applications. 

\textbf{Application Environment.}
IoT devices often ship with standard platforms and software pre-loaded so that potentially hundreds of devices do not have to be reconfigured across the wide area network. The VE should allow platforms and application environments to be pre-configured as part of the deployment, and the setup to be ready-to-use. Users should not be forced to individually configure each compute resources, though they should have the ability to do so if required.

\textbf{Scalable.}
IoT deployments can be large in the number of devices and sensors -- ranging in the 1000's -- and with complex network topologies. A VE should be able to scale to such large deployments with minimal resource and human overheads. At the same time, these devices offer real computing environments that require underlying compute capacities to be available on the host machine(s). Hence, we require the VE to \emph{weakly scale}, as long as the underlying infrastructure provides adequate cumulative compute and network capacity for all the devices. The use of elastic cloud resources as the host can enable this. 

\textbf{Reproducible.}
Simulators offer accurate reproducibility but limit the realism, or the ability to run real applications. Physical deployments are hard to get access to and may suffer from transient variability that affects reproducibility. A VE should offer a balance between running within a realistic deployment while being reproducible at a later point in time. This also allows easy sharing of deployment recipes for accurate comparisons.

\textbf{Cost effective.}
Clouds are able to offer a lower cost per compute unit due to economies of scale at data centers. But IoT devices while being commodity devices are costlier to purchase, deploy and manage. Having VEs offer comparable resource performance as the IoT deployment but for cheaper compute costs is essential. They should also make efficient use of the pay-as-you-go resources. Further, they should be deployable on-demand on elastic resources and release those resources after the experiments and validations are done.

\textbf{Ease of Design and Deployment.}
Users should be able to configure large IoT deployments with ease, and have them deploy automatically and rapidly. It should be possible to mimic realistic real-world topologies or generate synthetic ones for testing purposes.

\section{Architecture}
\label{sec:arch}

\begin{figure}[t!]
 \centering
	\subfloat[Architecture Design]{	
          \includegraphics[width=0.85\textwidth]{\gf{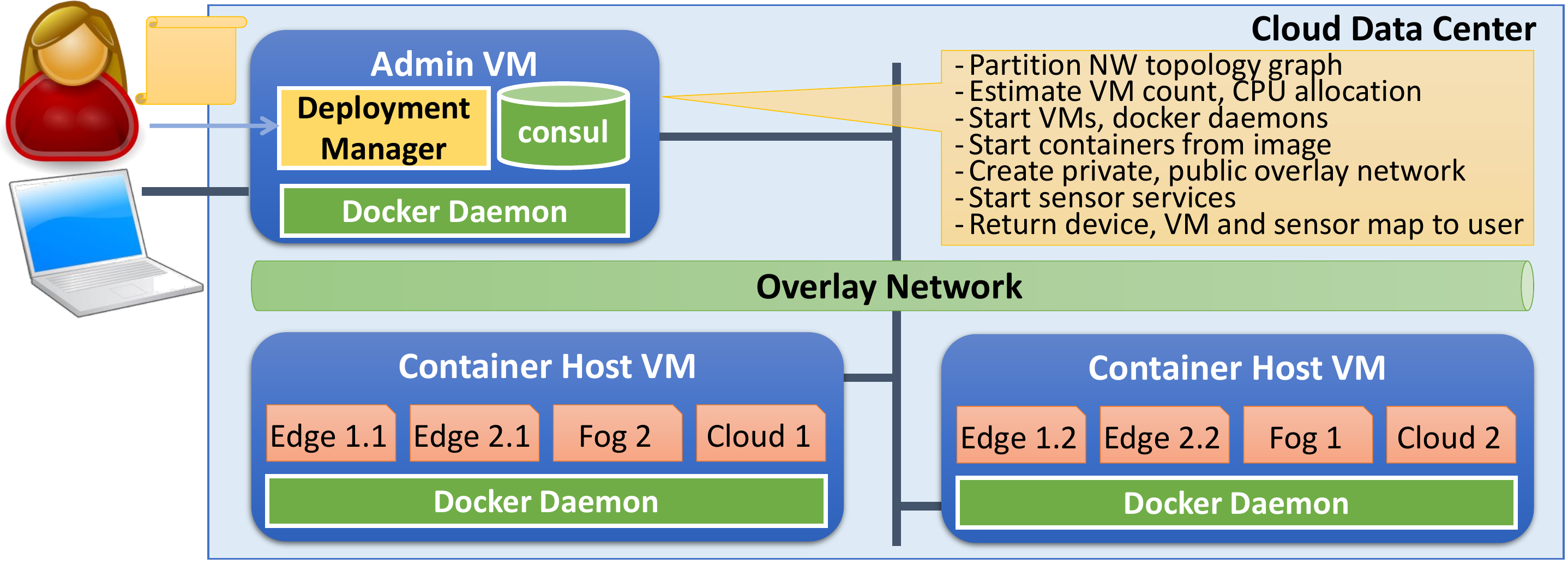}{arch.pdf}}
		\label{fig:arch}
	}\\
        \subfloat[JSON describing devices, sensors, VE deployment and host VMs.]{
          \includegraphics[width=1.0\textwidth]{\gf{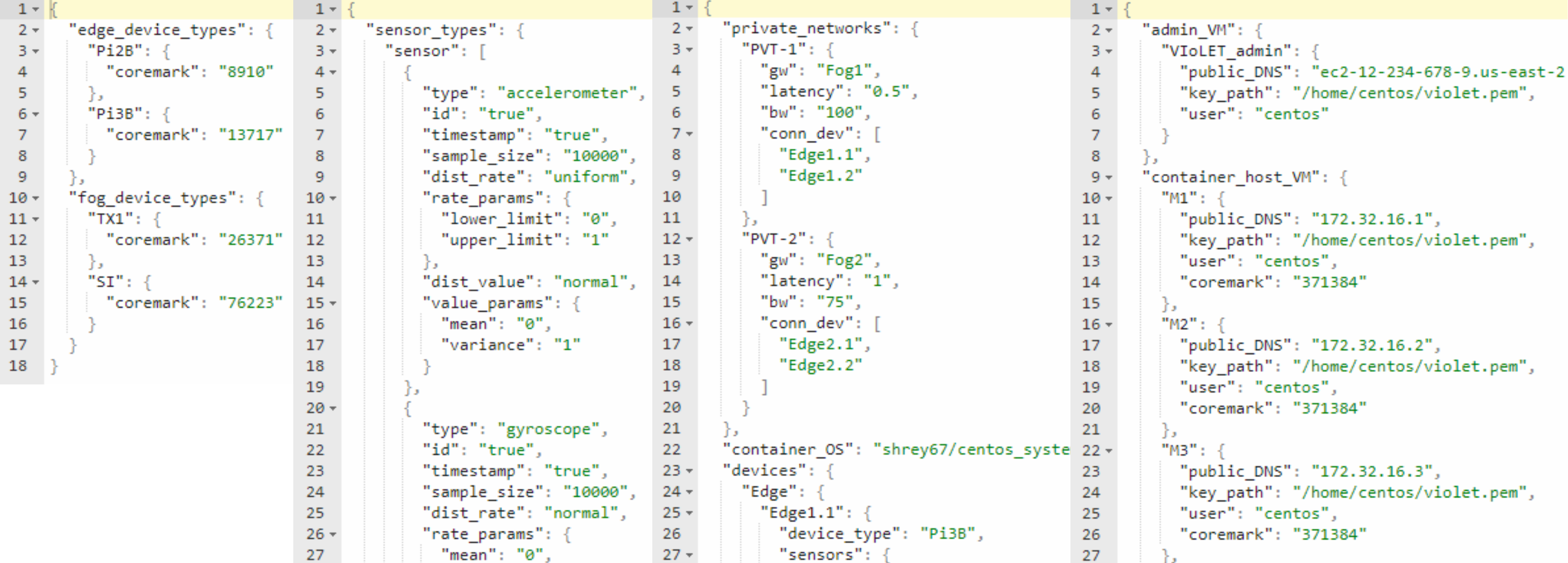}{json.pdf}}
	\label{fig:json}
      }

 \caption{\vlet Architecture and deployment documents}
\end{figure}

We give the high-level overview architecture of \vlet first, and then discuss individual components and design decisions subsequently. Fig.~\ref{fig:arch} shows the high-level architecture of our framework. Users provide their IoT VE as \emph{JSON deployment documents} (Fig.~\ref{fig:json}) that declaratively capture their requirements. A \texttt{devices.json} document lists the devices, their types (e.g., Raspberry Pi 3B, NVidia TX1) and their CPU performance.
Another, \texttt{sensors.json} document list the virtual sensors and their configurations available. Lastly, the actual deployment document, \texttt{deployment.json} lists the number of devices of various types, the network topology of the device inter-connects, including bandwidths and latencies, and optionally the virtual sensors and applications available on each device.

\vlet takes these documents and determines the number of cloud VMs of a specified type that are required to host containers with resources equivalent to the device types. It also decides the mapping from devices to VMs while meeting the compute capacity, and network bandwidth and latency needs of the topology, relative to what is made available by the host VMs.

Then, containers are configured and launched for each device using \emph{Docker}, and the containers are inter-connected through an 
overlay network. This allows different private and public networks to be created in the VE. Further, Traffic Control (TC) and Network Address Translation (NAT) rules are set in each container to ensure that the requested network topology, bandwidth and latency limits are enforced. 

Virtual sensors, if specified, are then started on each device and their streams available on a local network port in the container. Application environments or startup scripts if specified are also configured or launched. After this, the user is provided with a mapping from the logical device names in their deployment document to the physical device IPs of the matching container, and the VMs on which the containers are placed on. Users can access these devices using the Docker \texttt{exec} command. Further, the port numbers at which various logical sensors streams are available on each device is also reported back to the user. Together, these give access to the deployed runtime environment to the user.

\subsection{Compute Resources}
Containers are emerging as a light-weight alternative to VMs for multi-tenancy within a single host. They use Linux kernel's \texttt{cgroups} feature to offer benefits of custom software environment (beyond the OS kernel) and resource allocation and isolation, while having trivial overheads compared to hypervisors. They are well-suited for fine-grained resource partitioning and software sand-boxing among trusted applications.

Computing devices in \vlet are modeled as containers and managed using the \emph{Docker} automation framework. There are two parts to this: the \emph{resource allocation} and the \emph{software configuration}. Docker allows containers to have resource constraints to be specified~\footnote{Docker Resource Constraints, \href{https://docs.docker.com/config/containers/resource_constraints/}{docs.docker.com/config/containers/resource\_constraints}}. We use this to limit a container's capacity to match the CPU and Memory available on the native device. We use CPU benchmarks on the native device and the host machine to decide this allocation. The commonly used \emph{CoreMark}~\footnote{y, Embedded Microprocessor Benchmark Consortium (EEMBC), \href{http://coremark.org}{coremark.org}}
is currently supported for an integer-based workload, while \emph{Whetstone}~\footnote{Whetstone Benchmark History and Results, \href{http://www.roylongbottom.org.uk/whetstone.htm}{roylongbottom.org.uk/whetstone.htm}} has been attempted for floating-point operations. One subtlety is that while we use the multi-core benchmark rating of the device for the CPU scaling, this may map to fewer (faster) cores of the host machine.

A container's software environment is defined by the user as an image script (\texttt{Dockerfile}) that specify details like applications, startup services, and environment variables, and allow modular extensibility from other images. Public Docker repositories have existing images for common IoT platforms and applications (e.g., Eclipse Californium CoAP, Microsoft IoT Edge, RabbitMQ, Spark). \vlet provides a base image that includes its framework configuration and allow users to extend their device images from this base with custom software configuration. This is similar to specifying a VM image, except that the users are limited to the host device's Linux kernel OS~\footnote{Docker recently introduced support for Windows and Linux containers hosted on Windows Server using the Hyper-V hypervisor. But this is more heavy-weight than Linux containers, and not used by us currently.}. Hence, defining a compute device in \vlet requires associating a device type for resources, and a device image for the software environment.

\subsection{Network Topology}

\begin{figure}[t!]
 \centering
	\subfloat[Sample Topology Description]{
          \includegraphics[width=0.45\textwidth]{\gf{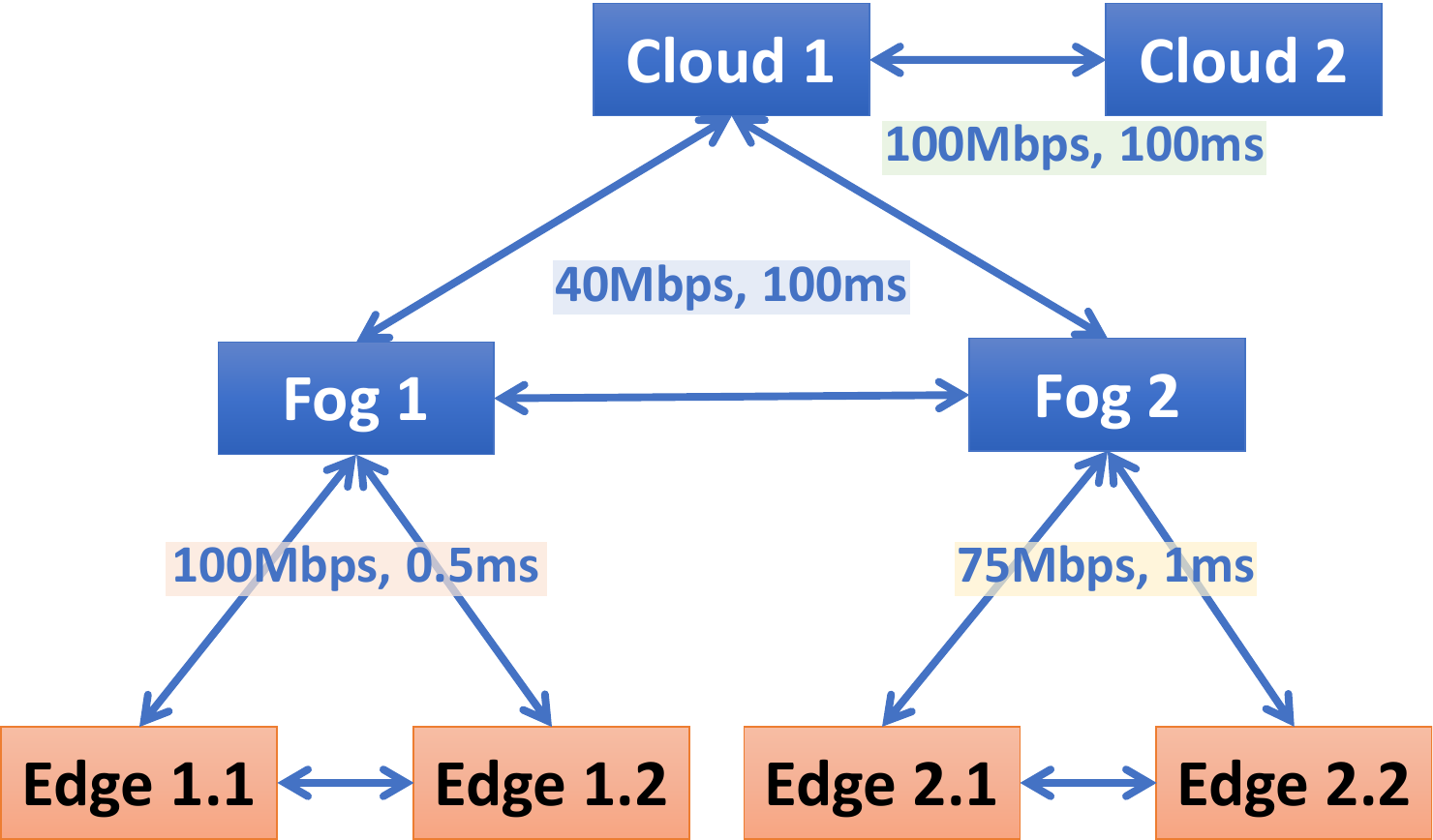}{nw-topo.pdf}}
	}
	\subfloat[Bridges in Overlay to Achieve Topology]{
~~~~~~          \includegraphics[width=0.4\textwidth]{\gf{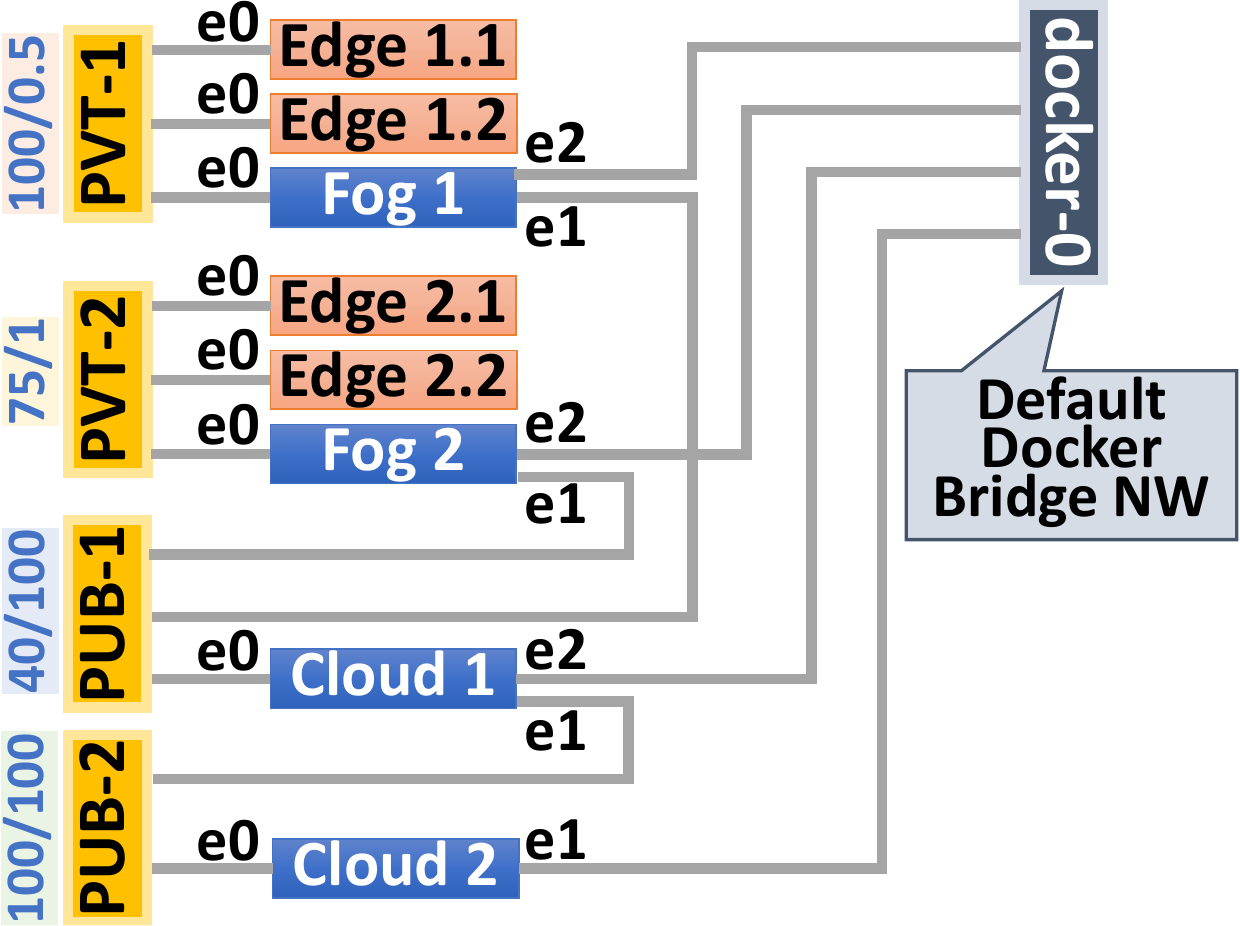}{nw-bridge.pdf}} ~~~~~~
	}
 \caption{Network Topology and Docker Overlay Network}
 \label{fig:nw}
\end{figure}

Users define the network topology for the devices based on three aspects: the public network or a private network the device is part of; the visibility of devices to each other as enforced by firewalls; and the bandwidth and latency between pairs of devices. IoT networks are usually composed of numerous private networks that interface with each other and the public Internet through gateways. We allow users to define logical private networks and assign devices to them. These exist in their own subnet. Each private network has a gateway device defined, and all traffic to the public network from other devices is routed through it. All gateway devices are part of one or more public networks, along with other devices that are on those public networks.

For simplicity, all devices in a private network by default can access each other, and have a common latency and bandwidth specified between pairs of devices by the user; and similarly for all devices connected to a public network. By default, devices on different public networks can reach each other. However, users can override this visibility between any pair of devices, and this is directional, i.e., $D1 \rightarrow D2$ need not imply $D1 \leftarrow D2$. 

We implement the bandwidth and latency between devices using \emph{Traffic Control (TC) rules} offered by Linux's \texttt{iproute2} utility, and the \texttt{network} service that we start on each container using \texttt{systemd}~\footnote{Traffic Control in Linux, \href{https://www.tldp.org/HOWTO/Traffic-Control-HOWTO/index.html}{tldp.org/HOWTO/Traffic-Control-HOWTO}}. Here, every unique bandwidth and latency requirement gets mapped to a unique virtual Ethernet port, and the rules and enforced on it. This Ethernet port is also connected to the bridge corresponding to the (private or public) network that the device belongs to. The bridges physically group devices that are on the same network, and also logically assign a shared bandwidth and latency to them. 
All devices on public networks are also connected to a common \texttt{docker-0} bridge for the VM they are present on, and which allows all to all communication by default.
Restricting the routing of traffic in a private network to/from the public network only through its gateway device is enacted through \texttt{ip} commands and \emph{Network Address Translation (NAT) rules}. These rules redirect packets from the Ethernet port connected to the private network, to the Ethernet port connected to the public network. 

Docker makes it easy to define connectivity rules and \emph{IP addressing} of containers present in a single host machine using custom bridges defined on the Docker daemon running on the host. However, devices in \vlet can be placed on disparate VMs and still be part of the same private network. Such communication between multiple Docker daemons requires custom Docker \emph{overlay networks}.
We create a standalone \emph{Docker Swarm pool} 
which gives us the flexibility to set network and system parameters~\footnote{Multi-host networking with standalone swarms, \href{https://docs.docker.com/network/overlay-standalone.swarm/}{docs.docker.com/network/overlay-standalone.swarm}}. For this, 
the host machines must be able to access a shared key-value store that maintains the overlay networking information. In \vlet, we use the \emph{Consul} discovery service as our key-value store that is hosted in a separate container on an \emph{admin VM}.

E.g., Fig.~\ref{fig:nw} shows a sample network topology, and the Ethernet ports and bridges to enact this in \vlet. Here, the edge devices E1.1, E1.2 form a private network PVT-1 with the fog device F1 as a gateway, and likewise E2.1, E2.2 and F2 form another private network, PVT-2. Each device can have sensors enabled to simulate data streams with different distributions. The bandwidth and latency within these private networks is uniform: 100Mbps/0.5ms for PVT-1, and 75Mbps/1ms for PVT-2. F1 and F2 fog devices go on to form a public network PUB-1 along with the cloud device, C1, with 40Mbps/100ms.
Similarly, the two cloud devices form another public network PUB-2, with 100Mbps/100ms. All these devices are on a single VM, and the public devices are also connected to the \texttt{docker-0} bridge for that VM.
While the edge devices are connected to a single overlay network, the fog and cloud devices can be connected to multiple overlay networks, based on bandwidth and latency requirements.

As can be seen, configuring the required network topology is complex and time consuming -- if done manually for each IoT deployment. Having a simple declarative document that captures the common network patterns in IoT deployments helps automate this.

\subsection{Sensors and Virtual Observation Streams}
Edge devices are frequently used to acquire IoT sensor data over hardware interfaces like serial, UART or I2C, and then make them available for applications to process and/or transfer. Experiments and validation of IoT deployments require access to such large-scale sensor data. To enable this, we allow users to define virtual sensors that are collocated with devices. These virtual sensors simulate the generation of sensed events and make them available at a local network port, which acts as a proxy for a hardware interface to the sensor. Applications can connect to this port, read observations and process them as required.

We support various configuration parameters for these sensors. The values for the sensor measurements themselves may be provided either as a text file with real data collected from the field, or as the properties of a statistical distribution, such as uniform random, Gaussian, and Poisson from which we sample and return synthetic values. In addition, the rate at which these values change or the events are generated is also specified by the user. 
Here too we can specify real relative timestamp or a distribution. 

We implement each sensor as a Python script that is launched as part of the container startup. The script starts a \emph{Flask} application server that listens on a local port. It takes the sensor's parameters, and internally starts generating observations corresponding to that. When a client connects to this port and requests a measurement, the service returns the \emph{current} reading. For simplicity, this is reported as a CSV string consisting of a user-defined logical sensor ID, the observation timestamp and a sensed value, but can be easily modified.

\subsection{Resource Mapping and Deployment}
The \emph{admin VM} runs a service that receives the user's deployment document as a REST request and enacts the deployment on cloud VMs in that data center. The default resource hosts are Amazon EC2 VMs  but this can easily be extended to resources on other cloud providers or even a private cluster. All AWS EC2 VM instances belong to a same Virtual Private Cloud (VPC) and the same subnet. On receipt of the deployment request, \vlet builds a graph of the network topology that is used to deploy the devices onto host resources. Here, the vertices of the graph are the devices and are labeled with the device's CPU requirement, given in the CPU benchmark metrics, e.g., iterations/sec for \emph{CoreMark}, and MWIPS for \emph{Whetstone}. An edge exists if a source device can connect to a sink device, and this is labeled by the bandwidth and latency for that network link. E.g., a private network where all devices can see each other will form a clique. 

We then make a gross estimate of the number of underlying resources we require. This is done by adding the vertex weights, dividing by the benchmark metric for the host (cloud VM) and rounding it up. This is the least number of identical host resources, say $n$, needed to meet the compute needs of all devices. 

Then, we partition the graph across these $n$ hosts using \texttt{gpmetis} such that the vertex weights are balanced across hosts and the sum of edge cuts between hosts, based on device bandwidths, is minimized.
This tries to collocate devices with high bandwidth inter-connects on the same host. We then check if the sum of the bandwidth edge cuts between devices in each pair of hosts is less than the available bandwidth capacity between them, and if the sum of benchmark metrics of all devices in a host is smaller than its capacity. 
If not, we increment $n$ by $1$ and repeat the partitioning, and so on.

This greedy approach provides the least number of host resources and the mapping that will meet the CPU and bandwidth capacities of the deployment. For now, we do not optimize for memory capacity and latency, but these can be extended based on standard multi-parameter optimization techniques.

\section{Evaluation}
\label{sec:eval}

We evaluate \vlet for two different IoT deployment configurations: \textbf{D105} with 105 edge and fog devices, and \textbf{D408} with 408 edge and fog devices. The configuration of each of the devices, their CoreMark CPU performance 
and the deployment counts are shown in Table~\ref{tbl:deploy}, along with the number of AWS VMs required to support them. CoreMark v1.0 is run with multi-threading enabled. 

\begin{table}[t]
\centering
\caption{Device Perf., Device Counts and Host VM Counts used in Deployments}
\begin{tabular}{l||r|r||r|r||r|r}
\hline
\multicolumn{3}{r||}{\emph{Deployment$\rightarrow$}}&\multicolumn{2}{c||}{\textbf{D105}} & \multicolumn{2}{c}{\textbf{D408}}\\
\hline
Device & Cores & CMark & Count & $\sum$CMark (k) & Count & $\sum$CMark (k)\\
\hline\hline
Pi 2B & 4 &   8,910   & 50 & 445& 200 & 1,782\\
Pi 3B & 4 &  13,717 & 50 & 685& 200 & 2,743\\
NVidia TX1 & 4 &  26,371  & 4	& 105 & 7 &184\\
Softiron & 8 &  76,223 & 1 & 76 &1 & 76\\
\hline
\hline
\multicolumn{3}{r||}{\emph{Total}}&\multicolumn{2}{r||}{\textbf{1,311}} & \multicolumn{2}{r}{\textbf{4,786}}\\
\hline
\hline
m4.10XL \emph{(host)} & 40 &  371,384 & 4 & 1,485 & 13 & 4,827 \\ 
\hline
\end{tabular}
\label{tbl:deploy}
\end{table}

We use two generations of \emph{Raspberry Pis} as edge devices -- \emph{Pi 2B} with $4 \times 900$~MHz ARM32 cores and \emph{Pi 3B} with $4\times1.2$~GHz ARM64 cores, and 1~GB RAM each. In addition, we have two fog resources -- a \emph{Softiron 3000 (SI)} with AMD A1100 CPU with $8 \times 2$~GHz ARM64 cores and 16~GB RAM, and an \emph{NVidia TX1} device with $4\times1.7$~GHz ARM64 cores and 4~GB RAM (its GPU is not exposed).
We use Amazon AWS \emph{m4.10XL} VMs that have $40 \times 2.4~GHz$ Intel Xeon E5-2676 cores, 160~GB RAM and 10~Gbps network bandwidth as the host. Each costs US\$2.00/hour in the US-East (Ohio) data center. As we see, the D105 deployment with $424$ ARM cores requires $3$ of these VMs with $120$ Xeon cores, and D408 with $1,636$ ARM cores requires $13$ of these VMs with $390$ Xeon cores. These deployments cost about US\$6/hour and US\$26/hour, respectively -- these are cheaper than a single Raspberry Pi device, on an hourly basis.

\subsection{Results for D105 and D408}
The network topology for these two deployments is generated synthetically. D105 is defined with 5 private networks and 4 public networks, while D408 has 8 private networks and 2 public networks. A fog device serves as the gateway in each private network, and we randomly place an equal number of edge devices in each private network. Their respective network configurations are given in Tables~\ref{tbl:d105:nw} and~\ref{tbl:d408:nw}. Each network has a fixed bandwidth and latency configuration, and this ranges from $5-100$~Mbps bandwidth, and $1-100$~ms latency, as specified. 
All devices in the public networks can see each other. Edge devices in the private network can access the public network, routed through their gateway, but devices in the public network cannot access the devices in the private network. 
It takes about 8~mins and 24~mins to launch these two topologies on \vlet.

\begin{table}[t]
\centering
\caption{Configuration of private and public networks in D105, and Deviation\% between Observed and Expected Bandwidth and Latency per network.}
\begin{tabular}{l||r|r||r|r||r|r}
\hline
\textbf{Network} & \multicolumn{2}{c||}{\textbf{Expected}}&\multicolumn{2}{c||}{\textbf{Obs. BW Dev.\%}}&\multicolumn{2}{c}{\textbf{Obs. Lat. Dev.\%}}\\
\cline{2-7}
& BW (Mbps) & Lat. (ms) & Median & Mean & Median & Mean \\
\hline\hline
PVT-1 & 5 & 25 & 11.0 & 11.0 & 0.6 & 0.5 \\
PVT-2 & 5 & 75 & 13.8 & 13.8 & 0.0 & 0.0 \\
PVT-3 & 25 & 1 & 4.8 & 4.8 & 15.0 & 15.5 \\
PVT-4 & 25 & 25 & 4.0 & 3.7 & 1.0 & 1.1 \\
PVT-5 & 25 & 50 & 1.6 & 1.4 & 0.0 & 0.0 \\
\hline
PUB-1 & 25 & 75 & -3.6 & -3.6 & 0.0 & 0.0 \\
PUB-2 & 25 & 75 & -3.6 & -3.6 & 0.0 & 0.0 \\
PUB-3 & 25 & 75 & -3.6 & -3.5 & 0.0 & 0.0 \\
PUB-4 & 25 & 75 & -3.6 & -3.6 & 0.0 & 0.0 \\
\hline
\end{tabular}
\label{tbl:d105:nw}
\end{table}

\begin{table}[t]
\centering
\caption{Configuration of private and public networks in D408, and Deviation\% between Observed and Expected Bandwidth and Latency per network.}
\begin{tabular}{l||r|r||r|r||r|r}
\hline
\textbf{Network} & \multicolumn{2}{c||}{\textbf{Expected}}&\multicolumn{2}{c||}{\textbf{Obs. BW Dev.\%}}&\multicolumn{2}{c}{\textbf{Obs. Lat. Dev.\%}}\\
\cline{2-7}
& BW (Mbps) & Lat. (ms) & Median & Mean & Median & Mean \\
\hline\hline
PVT-1 & 100 & 5 & -2.6 & -2.4 & 6.0 & 5.2 \\
PVT-2 & 75 & 5 & -1.1 & -1.3 & 3.0 & 4.9 \\
PVT-3 & 75 & 25 & -4.1 & -4.0 & 0.6 & 1.0 \\
PVT-4 & 50 & 5 & 0.0 & 0.1 & 4.0 & 4.9 \\
PVT-5 & 50 & 25 & -1.8 & -2.0 & 0.6 & 0.8 \\
PVT-6 & 25 & 25 & -1.8 & -2.0 & 0.6 & 0.8 \\
PVT-7 & 25 & 5 & 2.8 & 3.2 & 0.6 & 0.8 \\
PVT-8 & 25 & 50 & 4.8 & 5.0 & 0.6 & 0.8 \\ \hline
PUB-1 & 25 & 75 & -3.6 & -3.6 & 0.0 & 0.0 \\
PUB-2 & 25 & 100 & -7.0 & -7.0 & 0.0 & 0.0 \\
\hline
\end{tabular}
\label{tbl:d408:nw}
\end{table}

Once deployed, we run four baseline benchmarks to validate them. The first does \texttt{fping} between $2n$ pairs of devices in each private and public network, where $n$ is the number of devices in the network, and measures the observed latency on the defined links. Next, we sample a subset of $\frac{n}{2}$ links in each private and public network and run \texttt{iperf} on them to measure the observed bandwidth. Since \texttt{iperf} is costlier than \texttt{fping}, we limit ourselves to fewer samples. Third, we run \texttt{traceroute} to verify if the gateway device configured for each device matches the gateway of the private network, as a sanity check. These network sanity checks take $\approx 3~mins$ per network for D105, and run in parallel for all networks. Lastly, we run multi-core CoreMark concurrently on all devices.

\begin{figure*}[t]
	\centering
	\subfloat[D105 Deployment]{	
		\includegraphics[width=0.12\textwidth]{\gf{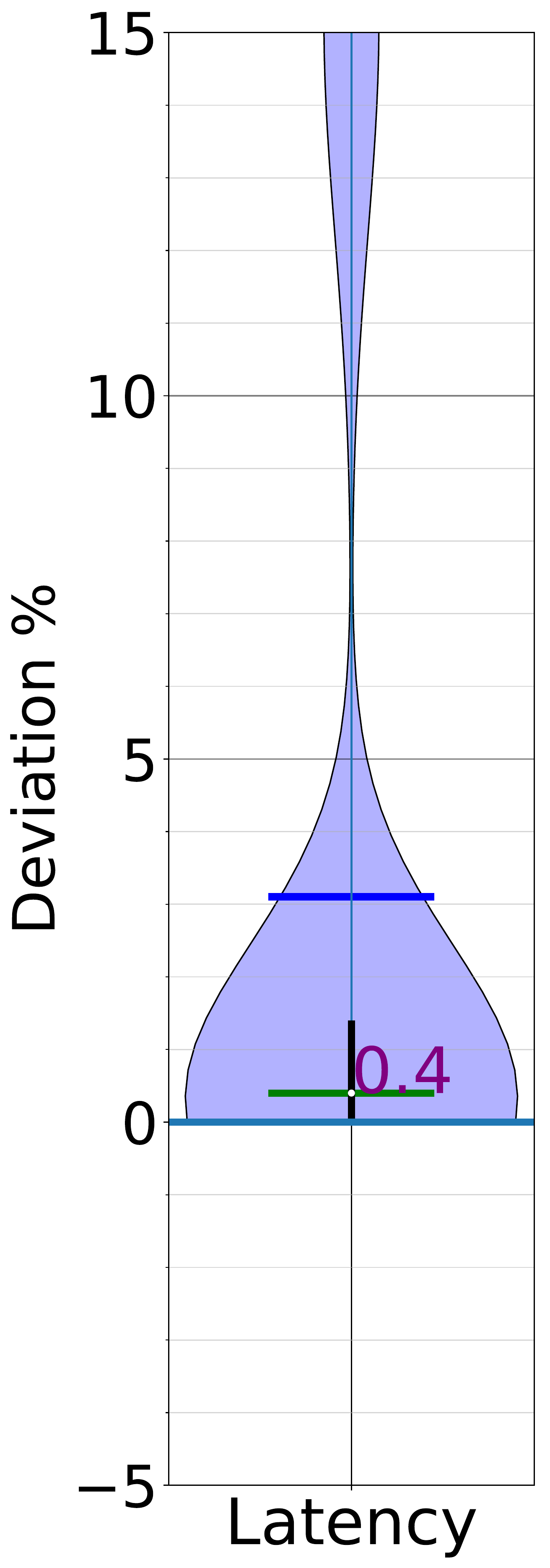}{d105-lat.pdf}}
		\includegraphics[width=0.12\textwidth]{\gf{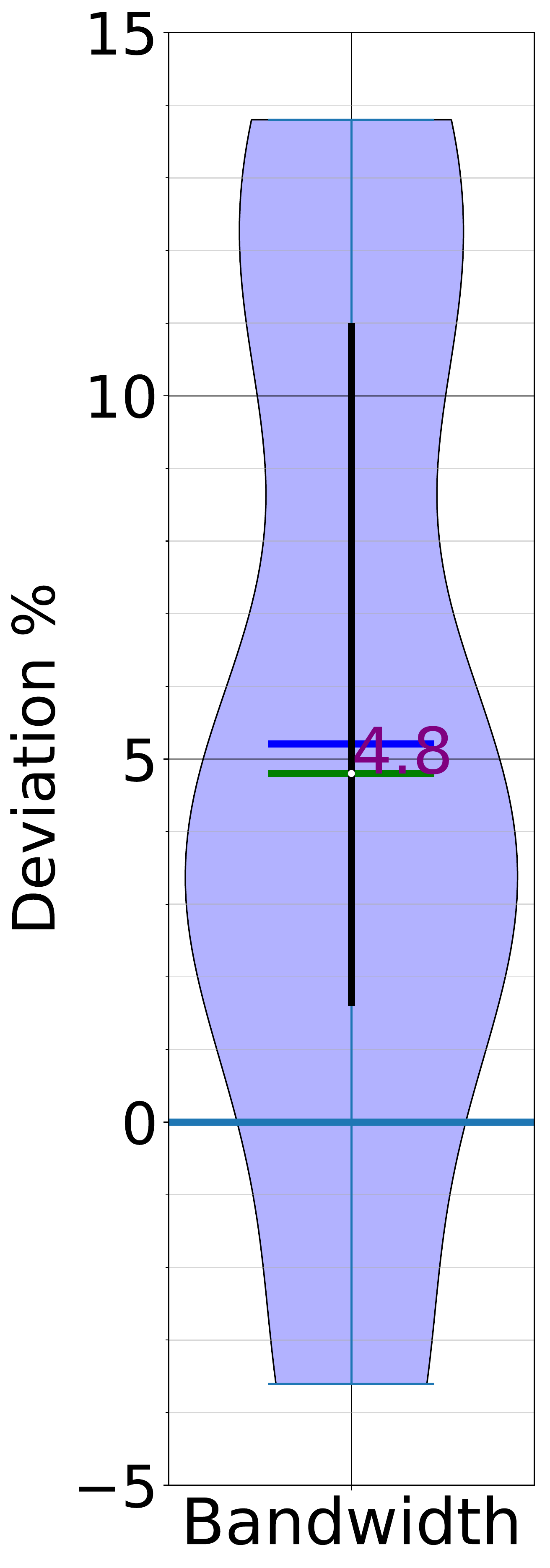}{d105-bw.pdf}}
		\includegraphics[width=0.21\textwidth]{\gf{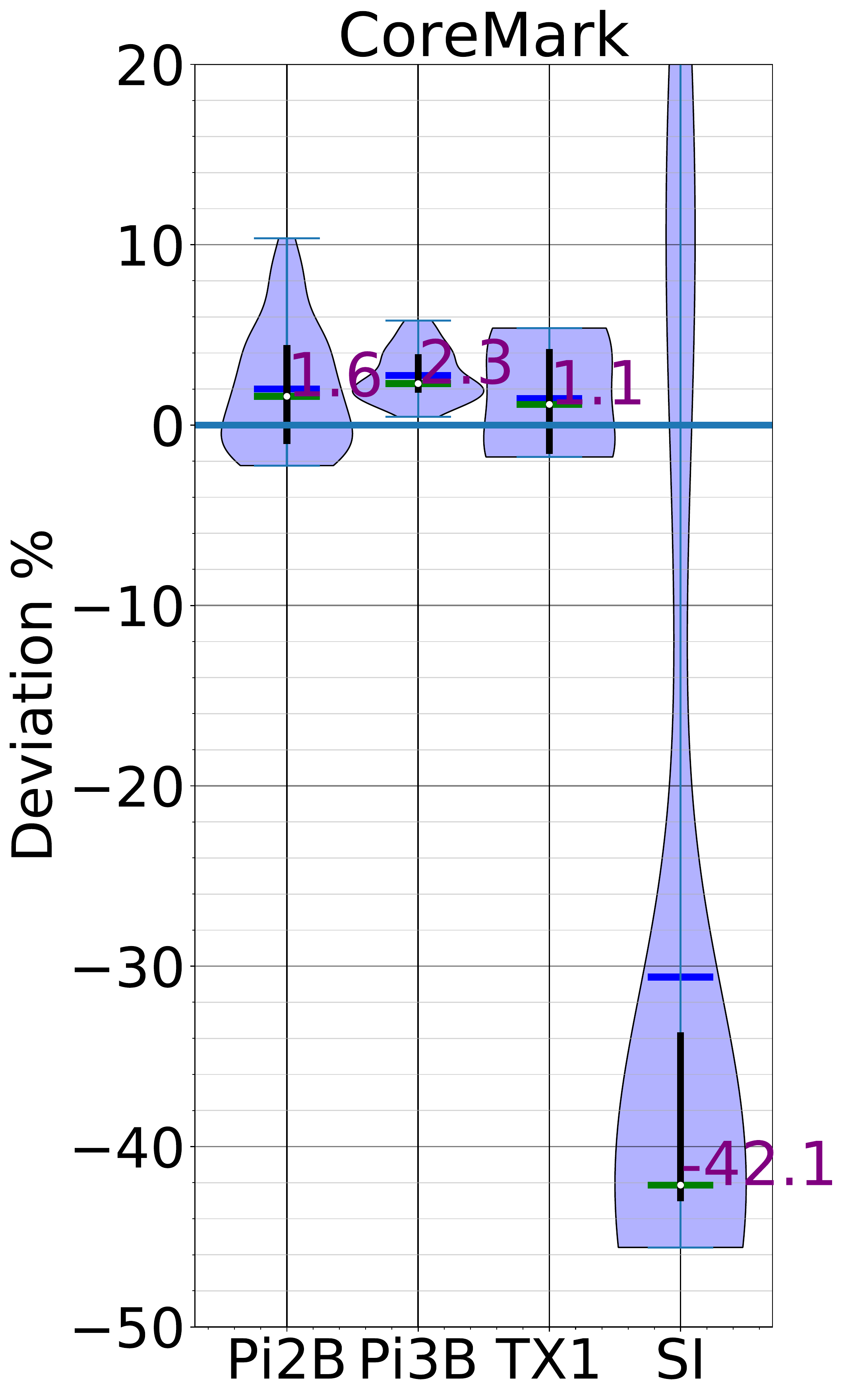}{d105-cpu.pdf}}
		\label{fig:perf:d105}
	}~~
        \subfloat[D408 Deployment]{
		\includegraphics[width=0.12\textwidth]{\gf{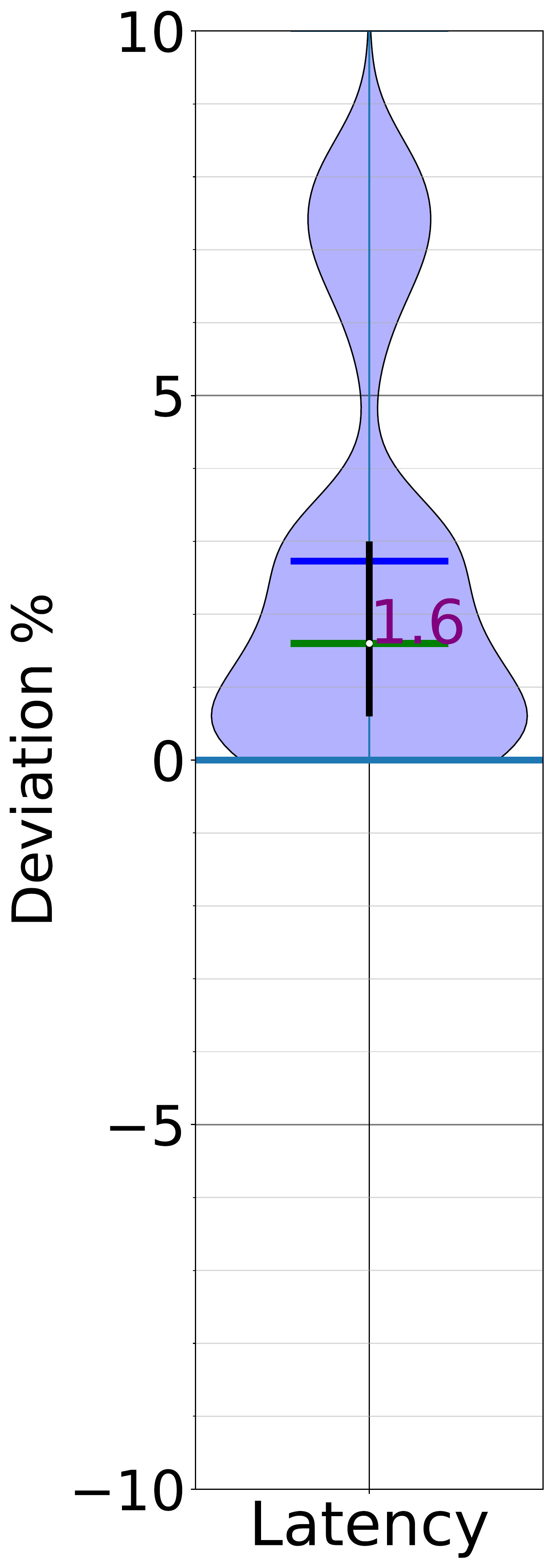}{d408-lat.pdf}}
		\includegraphics[width=0.12\textwidth]{\gf{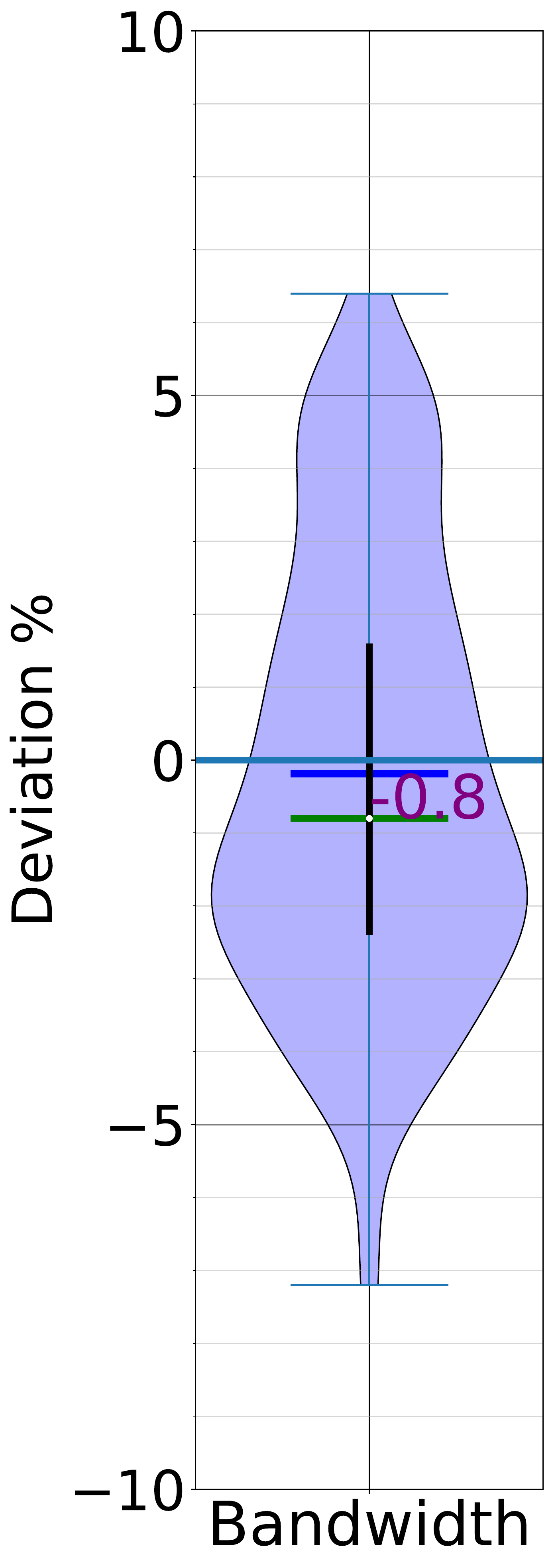}{d408-bw.pdf}}
		\includegraphics[width=0.21\textwidth]{\gf{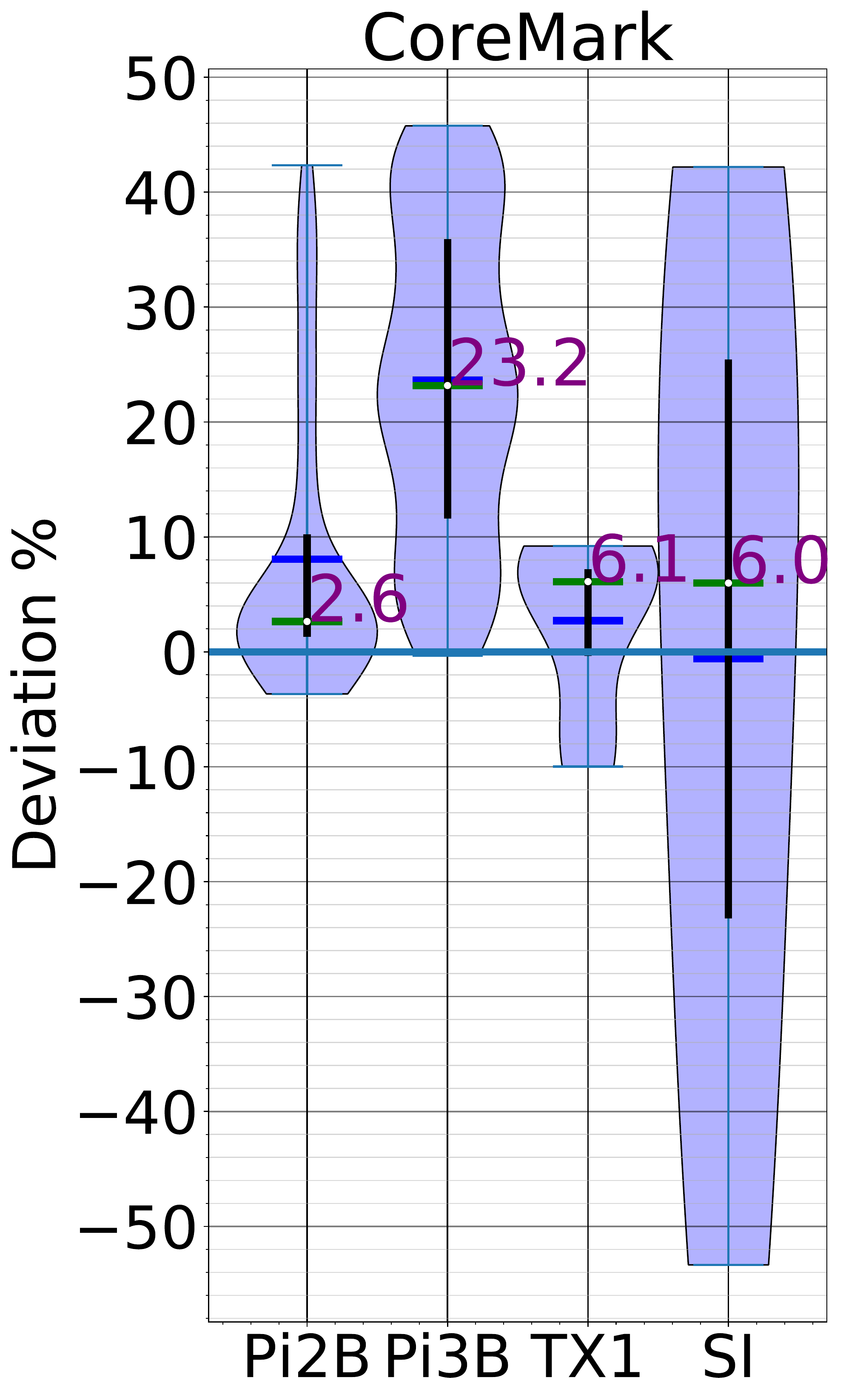}{d408-cpu.pdf}}
	\label{fig:perf:d510}
      }
\caption{Violin plot of \emph{deviation\%} for network latency, bandwidth and CoreMark CPU.}
\end{figure*}

Figs.~\ref{fig:perf:d105} and ~\ref{fig:perf:d510} show a violin plot of the \emph{deviation\%} of the observed network latency, bandwidth, and CoreMark performance from the expected metrics for the two deployments,  where $deviation\% = \frac{(Observed-Expected)}{Expected}\%$. The median value is noted in purple text.
We see that the median latency and bandwidth deviation\% are within $\pm 5\%$ for both the D105 and D408 deployments, with latency of $0.4\%$ and $1.6\%$, and bandwidth of $4.8\%$ and $-0.8\%$, respectively. This is within the margin of error for even real-world networks. 
The entire distribution in all these cases does not vary by more than $15\%$, showing a relatively tight grouping given the number of devices and VMs. 
We analyze these further for diverse network configurations in the next section.

We run the CoreMark CPU benchmark on all the devices concurrently and report the violin plot for the deviation\% for each of the $4$ device types. The median CoreMark value for each device is included in the violin, except for the SI fog where we report values from all the trials since there is just one such device in each deployment. 
We see that for the two Pis and TX1 -- the three slowest devices -- the median CoreMark deviation\% is within $\pm 2.5\%$ for D105, and the most deviation is $+10\%$ for Pi2B. These indicate that the observed performance is marginally higher than expected, and there is little negative deviation for these three devices. 
However, we see that the single SI fog device, which is the largest device, has a median deviation\% of $-42.1\%$ from 40 trials of CoreMark that were run on it. The distribution is also wide, ranging from $-45\%$ to $+21\%$. This indicates that the concurrent multi-threaded CoreMark runs on 10's of containers on the same VM is causing the largest device container to have variable performance.
In fact, the sum of the observed CoreMarks for all the deployed devices in D105 is $1,319k$, which is close to the sum of the expected CoreMark from the devices of $1,311k$. So the small over-performance of many small devices is causing the under-performance of the large device.
D408 shows a different behavior, with Pi3B showing higher positive deviations, with a median of $23.2\%$, while the other devices show a smaller positive deviation of $2.6$--$6\%$. SI however does show a wider distribution of the deviation\% as before.

Besides these baseline network and CPU metrics, we also run two types of application workloads. One of them starts either an MQTT publisher or a subscriber on each device, and each connects to an \emph{Eclipse Mosquitto MQTT broker} on its gateway. A publisher samples observations from a local sensor and publishes it to a unique topic at its gateway broker while a subscriber subscribes to it. This tests the network and process behavior for the common pub-sub  pattern seen in IoT. While results are not plotted due to lack of space, we observe that the median end-to-end latency for each message is $\approx 50~ms$, which loosely corresponds to the two network hops required from the publisher to the broker, and broker to subscriber.

Another workload that we evaluate is with the ECHO dataflow platform for edge and cloud~\cite{echo}. Here, we incrementally launch 100 Extract-Transform-Load dataflows using the Apache NiFi engine on distributed devices and observe the latency time for deployment and the end to end latency for the dataflows. This is yet another use-case for \vlet to help evaluate the efficacy of such edge, fog and cloud orchestration platforms and schedulers.

\subsection{Analysis of Network Behavior}
\begin{figure*}[t]
	\centering
        \subfloat[Latency]{
		\includegraphics[page=2,width=0.4\textwidth]{\gf{plots/camera/nw-plots.pdf}{nw-plots.pdf}}
	\label{fig:perf:nw-lat}
      }~~
	\subfloat[Bandwidth, at different Latencies. Bottom row shows ideal bandwidth for latency.]{	
		\includegraphics[page=1,width=0.5\textwidth]{\gf{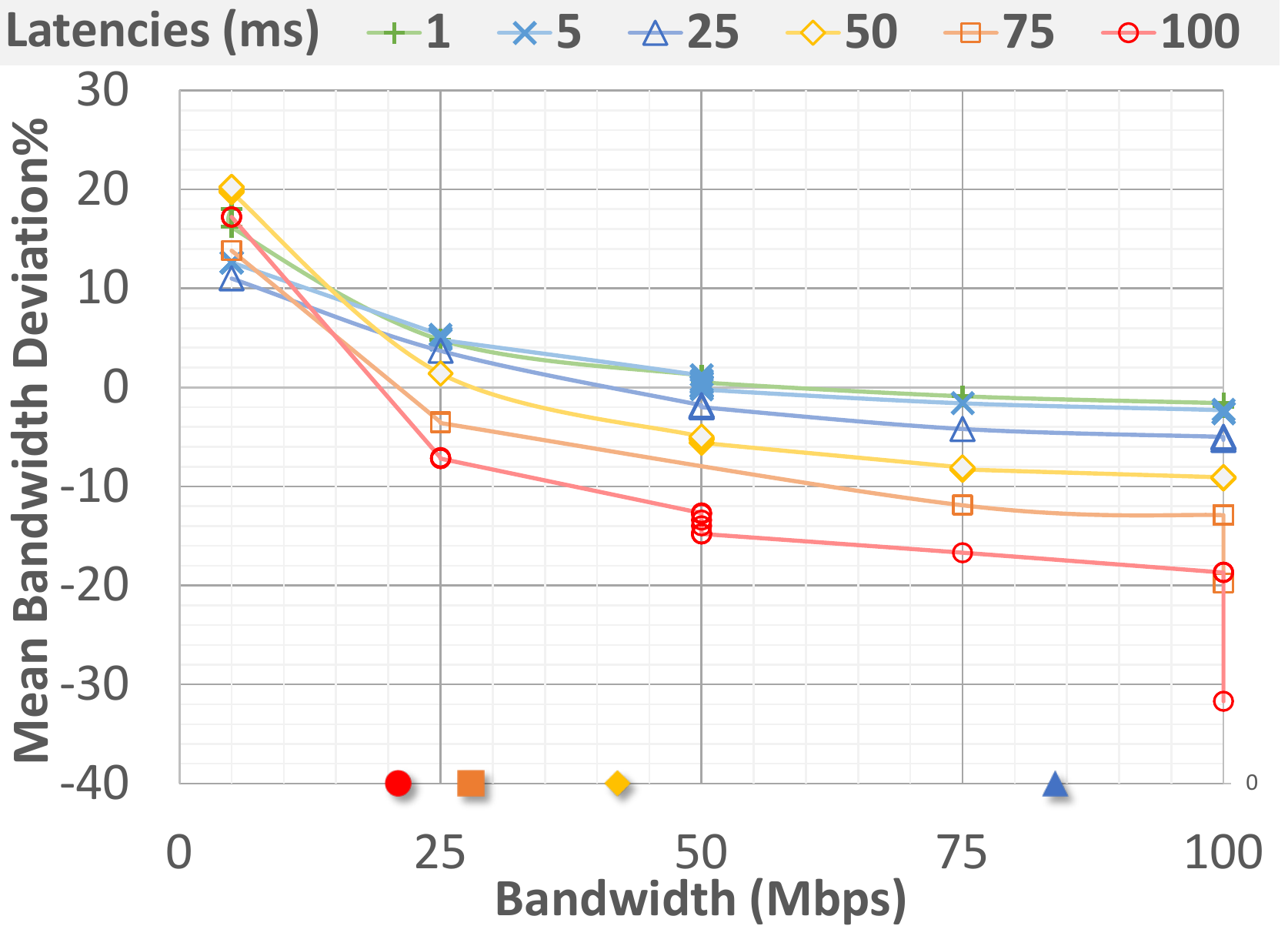}{nw-plots.pdf}}
		\label{fig:perf:nw-bw}
	}
\caption{Variation of \emph{deviation\%} for different latency and bandwidth configurations.}
\label{fig:perf:nw}
\end{figure*}

Being able to accurately model network behavior is essential for IoT VEs. Here, we perform more detailed experiments that evaluate the impact of specific bandwidth and latency values on the deviation\%. Specifically, we try out 19 different network configurations of the D105 deployment while varying the pair of bandwidth and latency values on these networks. These together form 143 different networks. In Fig.~\ref{fig:perf:nw-bw}, we plot the deviation\% of the mean bandwidth, as the bandwidth increases for different latency values, while in Fig.~\ref{fig:perf:nw-lat}, we plot the deviation\% of the mean latency, as latency increases.

It is clear from Fig.~\ref{fig:perf:nw-lat} that the latency deviation is sensitive to the absolute latency value. For small latency values of $1$~ms, the deviation\% ranges between $15-40\%$, and this drops to $2.6-8\%$ for $5$~ms. The deviation\% exponentially reduces for latencies higher than that, with latencies over $50$~ms having just $0.1\%$ deviation. The latency between VMs is measured at $0.4$~ms, while between containers on the same VM is $0.06$~ms. Hence, achieving a latency better these is not possible, and the achieved latency depends on the placement of containers on the same or different VMs.
Since our network partitioning currently is based on bandwidth and compute capacity, and not latency limits, it is possible that two devices requiring low latency are on different VMs. As a result, the deviation\% increases. Here, we see that the latency deviation is independent of the bandwidth of the network link.

We observe that the deviation in bandwidth is a function of both latency and bandwidth. In fact, it is also a function of the \emph{TCP window size}, which by default is set to $262,144$~bytes in the containers. The \emph{Bandwidth Delay Product (BDP)} is defined as the product of the bandwidth and latency. For efficient use of the network link, the TCP window size should be greater than this BDP, i.e., $Window \ge Bandwidth \times Latency$. In other words, given a fixed latency and TCP window size, the $Peak~Bandwidth = \frac{Window}{Latency}$.

Fig.~\ref{fig:perf:nw-bw} shows the bandwidth deviation\% on the Y axis for different latencies, as the bandwidth increases on the X axis. It also shows the maximum possible bandwidth for a given latency (based on the window size) along the bottom X axis. We observe that for low latencies of $1$ -- $25$~ms, the bandwidth deviation\% is low and falls between $-5.1$ -- $18\%$ for all bandwidths from $5$ -- $100$~Mbps. This is because with the default window size, even a latency of $25$~ms supports a bandwidth of $83$~Mbps, and lower latencies support an even higher peak bandwidth. The positive deviation\% is also high for low bandwidth values and lower for high bandwidth values -- even small changes in absolute bandwidth causes a larger change in the relative deviation\% when the bandwidth is low. 

We also see that as the latency increases, the negative deviation\% increases as the bandwidth increases. In particular, as we cross the peak bandwidth value on the X axis, the deviation\% becomes more negative. E.g., at $75$~ms, the peak bandwidth supported is only $28$~Mbps, and we see the bandwidth deviation\% for this latency worsen from $-3.6\%$ to $-11.9\%$ when the bandwidth configuration increases from $25$~Mbps to $75$~Mbps. This is as expected, and indicates that the users of the container need to tune the TCP window size in the container to enforce bandwidths more accurately. 

\section{Related Work}
\label{sec:related}

The growing interest in IoT and edge/fog computing has given rise to several \emph{simulation environments}. \emph{iFogSim}~\cite{gupta:2016} extends the prior work on CloudSim~\cite{Calheiros:2011} to simulate the behavior of applications over fog devices, sensors and actuators that are connected by a network topology. Users define the compute, network and energy profiles of fog devices, and the properties and distributions of tuples from sensors. DAG-based applications with tasks consuming compute capacity and bandwidth can be defined by the user, and its execution over the fog network is simulated using an extensible resource manager. The goal is to evaluate different scheduling strategies synthetically. We similarly let devices, network and sensors to be defined, but actually instantiate the first two -- only the sensor stream is simulated. This allows users to evaluate real applications and schedulers.

\emph{Edgecloudsim}~\cite{sonmez:2017} offers similar capabilities, but also introduces mobility models for the edge into the mix. They simulate network characteristics like transmission delay for LAN and WAN, and also task failures due to mobility for a single use-case. \emph{IOTSim}, despite its name, simulates the execution of Map Reduce and stream processing tasks on top of a cloud data center, and uses CloudSim as the base simulation engine. While IoT motivates the synthetic application workloads for their big data platform simulation, they do not actually simulate an IoT deployment.

In the commercial space, city-scale simulators for IoT deployments in smart cities are available~\cite{leland:2017}. These mimic the behavior of not just devices, sensors, actuators and the network, but also application services like MQTT broker and CoAP services that may be hosted. These offer a comprehensive simulation environment for city-planners to perform what-if analysis on the models. We go a step further and allow realistic devices and networks to be virtualized on elastic cloud VMs, and applications themselves to be executed, without physically deploying the field devices.
Simulators are popular in other domains as well, such as cloud, network and SDN simulators~\cite{Calheiros:2011,henderson:2006,mininet}.

There have been container-based solutions that are closer to our approach, and allow large-scale customized environments to be launched and applications to be run on them. Ceesay, et al.~\cite{ceesay:2017}, deploy container-based environments for Big Data platforms and workloads to test different benchmarks, ease deployment and reduce reporting costs. Others have also used such container-based approaches to inject faults into the containers, and evaluate the behavior of platforms and applications running on them~\cite{jdd:2016}.

Other have proposed IoT data stream and application workloads for evaluating big data platforms, particularly stream processing ones. Here, the sensor data is simulated at large-scales while maintaining realistic distributions~\cite{gu:2015,arlitt:2015}. These can be used in place of the synthetic sensor streams that we provide. Our prior work has proposed stream and stream processing application workloads for IoT domains~\cite{shukla:cpe:2017}. These can potentially use \vlet for evaluating execution on edge and fog, besides just cloud resources.

\emph{Google's Kubernetes~\cite{kuber}} is a multi-node orchestration platform 
for container life-cycle management. It schedules containers across nodes to balance the load, but 
is not aware of network topologies that are overlaid on the containers. 
\vlet uses a simple graph-partitioning approach for placement of containers on VMs to balance the CPU capacity, as measure by CoreMark, and ensure that the required device bandwidths stay within bandwidth available between the hosts.

\section{Conclusions and Future Work}
\label{sec:conclusions}
In this paper, we have proposed the design requirements for a Virtual IoT Environment, and presented \vlet to meet these needs. \vlet allows users to declaratively create virtual edge, fog and cloud devices as containers that are connected through user-defined network topologies, and can run real IoT platforms and applications. This offers first-hand knowledge of the performance, scalability and metrics for the user's applications or scheduling algorithms, similar to a real IoT deployment, and at large-scales. It is as simple to deploy and run as a simulation environment, balancing ease and flexibility, with realism and reproducibility on-demand. It is also affordable, costing just US\$26/hour to simulate over $400$ devices on Amazon AWS Cloud. \vlet serves as an essential tool for IoT researchers to validate their outcomes, and for IoT managers to virtually test various software stacks and network deployment models.

There are several extensions possible to this initial version of \vlet.
One of our limitations is that only devices for which container environments can be launched by Docker are feasible. While any device container that runs a standard Linux kernel using \texttt{cgroups} (or even a Windows device~\footnote{Docker for Windows, https://docs.docker.com/docker-for-windows/})) can be run, this limits the use of edge micro-controllers like Arduino, or wireless IoT motes that run real-time OS.
Also, leveraging Docker's support for GPUs in future will help users make use of accelerators present in devices like NVidia TX1~\footnote{GPU-enabled Docker Containers, https://github.com/NVIDIA/nvidia-docker}.
There is also the opportunity to pack containers more efficiently to reduce the cloud costs~\cite{awada:2017}, including over-packing when devices will not be pushed to their full utilization.

Our network configurations focus on the visibility of public and private networks, and the bandwidth and latency of the links. However, it does not yet handle more fine-grained transport characteristics such as collision and packet loss that are present in wireless networks. Introducing variability in bandwidth, latency, link failures, and even CPU dynamism is part of future work.
More rigorous evaluation using city-scale models and IoT applications are also planned using large private clusters to evaluate \vlet's weak scaling.

\section{Acknowledgments}
This work is supported by research grants from VMWare, MHRD and Cargill, and by cloud credits from Amazon AWS and Microsoft Azure. We also thank other DREAM:Lab members, Aakash Khochare and Abhilash Sharma, for design discussions and assistance with experiments. We also thank the reviewers of Euro-Par for their detailed comments that has helped us improve the quality of this paper.

%
%
\bibliographystyle{plain}
\bibliography{paper}


\end{document}